\documentclass[journal]{IEEEtran}
\usepackage{amsmath}
\usepackage{amssymb}
\usepackage{amsthm}
\usepackage{cite}
\usepackage{color}
\usepackage{xcolor}
\usepackage{caption}
\usepackage{epsfig,latexsym}
\usepackage{float}
\usepackage{fancyhdr}
\usepackage{graphicx}
\usepackage{indentfirst}
\usepackage{mdwmath}
\usepackage{mdwtab}
\usepackage{subfigure}
\usepackage{setspace}
\usepackage{times}
\usepackage{url}
\usepackage{multirow}
\usepackage{algorithm}
\usepackage{algorithmic}
\usepackage{array}
\usepackage{tabularx}

\newtheorem{theorem}{Theorem}

\newtheorem{lemma}{Lemma}

\newtheorem{corollary}{Corollary}

\begin{document}

\title{Rethinking Satellite Networks: When Navigation Meets Communications}
\author{Tianwei Hou, \emph{Member, IEEE}, Da Guan, Yu Zhang, Anna Li, \emph{Member, IEEE},\\ and Arumugam Nallanathan, \emph{Fellow, IEEE}

\thanks{This work is supported in part by the Beijing Natural Science Foundation L232041, and in part by the EPSRC grant numbers to acknowledge are EP/W004100/1, EP/W034786/1 and EP/Y037243/1,. (Corresponding author: )}
\thanks{T. Hou, D. Guan, Y. Zhang are with the School of Electronic and Information Engineering, Beijing Jiaotong University, Beijing 100044. (email: twhou@bjtu.edu.cn; 23111014@bjtu.edu.cn; 25110072@bjtu.edu.cn).}
\thanks{A. Li is with the School of Computing and Communications, Lancaster University, Lancaster LA1 4WA, U.K. (e-mail: a.li16@lancaster.ac.uk).}
\thanks{A. Nallanathan is with the School of Electronic Engineering and Computer Science, Queen Mary University of London, London E1 4NS, U.K., and also with the Department of Electronic Engineering, Kyung Hee University, Yongin-si, Gyeonggi-do 17104, Korea. (e-mail: a.nallanathan@qmul.ac.uk).}
}

\maketitle

\begin{abstract}
This paper investigates satellite navigation and communication systems in both low-Earth-orbit (LEO) and medium-Earth-orbit (MEO) satellites, which systematically outlines the fundamental principles of satellite navigation systems (SNS), satellite communication systems (SCS), and integrated navigation and communication (INAC) systems. By exploring the enhanced capabilities of satellite systems, the article emphasizes how INAC systems improve overall functionality by enabling efficient signal multiplexing and multiple access, positioning multi-functional satellites as promising alternatives to traditional architectures. Moreover, it introduces emerging frontiers for LEO-based SNS and MEO-based SCS through the integration of advanced sixth-generation (6G) wireless technologies, which cannot be realized through mere extensions of existing communication or navigation techniques. Motivated by these insights, the article further discusses various conceptual transitions required to unlock the full potential of INAC systems, with particular focus on channel capacity, positioning accuracy, and artificial intelligence-enabled waveform design.
\end{abstract}

\section{Introduction}
Satellite navigation and communication technologies are gaining attention as promising solutions for the next generation of wireless networks. In current mainstream satellite system deployments, the satellite navigation system (SNS) offers high-precision positioning, navigation, and timing (PNT) services through both medium-Earth-orbit (MEO) and Geostationary-orbit (GEO) satellites, catering to air, terrestrial, and marine users. Meanwhile, the satellite communication system (SCS), leveraging low-Earth-orbit (LEO) satellites, provides supplementary services, especially in rural and marine areas, for sixth-generation (6G) wireless networks~\cite{ref1}. However, satellite navigation and communication systems are typically deployed independently based on their specific requirements and geometric distributions, which raises a natural question: Can we integrate both navigation and communication capabilities into each satellite to deliver comprehensive PNT and communication services?

Integrating multiple functions into satellites operating on different orbits presents several challenges, as outlined below: i) Due to significant free-space path loss over long distances, the received signal-to-noise ratio (SNR) of MEO satellites drops below 0 dB. Without additional antenna gain or transmit power, such satellites are mainly suitable for navigation services. ii) While LEO satellites can offer communication services to ground users, their geometric distribution and time of passing are not conducive to navigation compared to MEO. However, with the advent of advanced 6G technologies, it becomes feasible to combine both satellite navigation and communication into a single satellite~\cite{ref2}. By treating communication and navigation as distinct functions with separate resource blocks, a virtuous cycle is initiated, fostering cross-functional synergies.

In specific scenarios, especially when satellites within traditional global navigation satellite system (GNSS), such as Beidou and Galileo SNS, fail to provide positioning services to users, the integrated navigation and communication (INAC) system may become the preferred solution for other potential novel services tailored to diverse application demands in 6G. While the INAC increases the complexity of satellite systems, it also creates new possibilities for personalized navigation and communication services. Additionally, the fusion of INAC facilitates functions that would be impossible by merely extending line-of-sight (LoS) links in the blocked environment. The paradigm shift enabled by INAC introduces both advantages and challenges, prompting a reevaluation of INAC system design. These serve as the primary motivations for this article.

The primary contributions presented in this paper are summarized below:
\begin{itemize}
  \item A systematic review of the core characteristics of SNS and SCS in both LEO and MEO satellite systems.
  \item A thorough comparative analysis of the similarities and distinctions between SNS and SCS in MEO and LEO satellite systems, focusing on: i) the deployment of SNS within the LEO satellite system, and ii) the necessary infrastructure for the MEO SCS.
  \item A rethinking of the evolution towards INAC systems, with particular emphasis on channel capacity, positioning accuracy, and artificial intelligence (AI)-enabled waveform design.
\end{itemize}

Unlike existing studies that focus on individual techniques, this paper establishes a unified framework for INAC satellite networks, revealing the paradigm shifts from separate systems to integrated design. This unified framework synergizes LEO/MEO orbital advantages with RIS-assisted technologies. It provides a systematic design logic for INAC that has not been proposed in prior LEO-PNT or RIS-related works.

\section{Fundamentals of Satellite Navigation and Communication}
%From independent to integrated
%One system figure and two equations
%at most 1 figure, 1000 words
%Describe the difference between navigation and communication, LEO and MEO

\begin{figure}[ht]
\centering
\includegraphics[width =\linewidth]{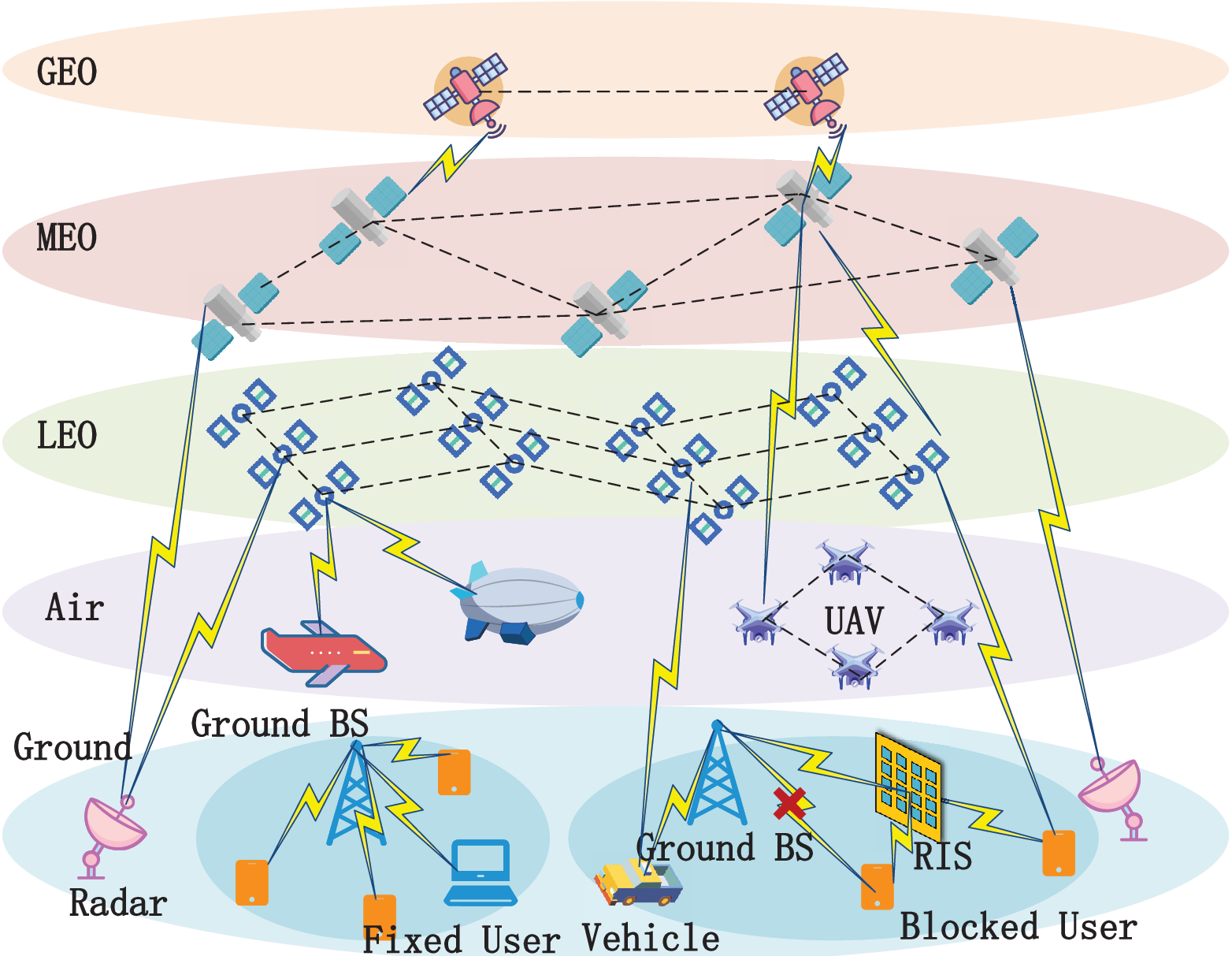}
\caption{ \centering SNAC systems.}
\label{sys}
\end{figure}

Different from traditional terrestrial systems which mostly rely on fixed infrastructures, satellite navigation and communication systems leverage space, aerial, and ground layers to provide global coverage and enhanced services with a comprehensive depiction of the system architecture. The satellite navigation and communication (SNAC) systems are shown in Fig.~\ref{sys}.

\renewcommand{\arraystretch}{1.0}
\setlength\tabcolsep{2.0pt}%调列距

\subsection{Fundamentals of SCS }\label{FunSC}%350-400 369

The traditional SCS architecture works together to achieve global coverage and data transmission, which comprises three main components: the space segment, the ground segment, and the user segment. Multi-tier satellites are distributed at different orbital altitudes in the space segment. The ground segment is responsible for communicating and controlling the satellites, while the user segment is composed of various user terminals~\cite{nav}.

The communication distance from a satellite to the ground is usually much greater than that of the ground-based nodes, no matter the orbits of the satellite employed. Orbiting within the altitude range of 160 to 2000 km, LEO satellites exhibit constrained coverage areas. However, their proximity to Earth results in low signal propagation delays, typically ranging from a few milliseconds to several tens of milliseconds. The low latency makes LEO satellites suitable to support real-time communication and high-precision positioning applications. Additionally, LEO satellites can achieve higher data rates, often one to two orders of magnitude greater than those of MEO satellites, with orbital altitudes ranging from 2000 to 35786 km. MEO satellites provide extensive coverage areas, which allows fewer satellites to provide wide-area services. However, the higher altitude results in higher signal propagation delays, typically on the order of several hundred milliseconds.

\subsection{Fundamentals of SNS}%350-400
%Satellite navigation is realized by the collaborative operation of the space segment, ground segment, and user segment.

The SNS is realized by the collaborative operation of the space segment, ground segment, and user segment~\cite{Hein2020}. The space segment includes LEO, MEO, and GEO satellites orbiting at varying altitudes with optimized trajectories to ensure global coverage. Each satellite carries atomic clocks synchronized via ground stations and inter-satellite links (ISLs). Satellites broadcast navigation signals on multiple carrier frequencies, comprising pseudo-range codes and carrier phase that carry ephemeris data, clock corrections, atmospheric model parameters, satellite health information, and almanac data. Receivers process navigation messages to determine satellite positions, correct signal delays, and assess satellite availability, enabling accurate and reliable global positioning.

The ground segment consists of global networks of monitoring stations and master control centers. Monitoring stations collect satellite signals in real time, while providing atmospheric data used to generate delay correction models.
The master control center calculates precise ephemeris and satellite clock corrections based on collected data and atmospheric models.

The receiver in user segment determines position by solving geometric relationships based on signals from at least four satellites. The pseudo-range measurement estimates signal travel time through code correlation, providing initial distance information. Carrier phase measurement tracks the phase of the carrier wave, offering higher precision for positioning. Satellite signals undergo corrections for ephemeris, satellite clock errors, and atmospheric delays, and then the corrected observations are processed by using the least square method (LSM) and Kalman filtering methods to estimate the receiver’s three-dimensional position, velocity and receiver clock bias.

In SNS, the LEO constellation and the MEO constellation are different and complementary.
LEO satellites deliver signals more than 30 dB stronger than those from MEO satellites while simultaneously offering diverse elevation angles that improve dilution of precision in urban canyons~\cite{Xu2025}.
Kilohertz-scale Doppler shifts enable millimetre-level time transfer alongside two-way ranging yet oblige rapid frequency search together with predictive ephemerides, because LEO satellites sweep across the sky within minutes. By contrast, MEO constellations offer geometrically stable visibility, while their highly stable clocks facilitate reliable stand-alone single-system positioning. A forthcoming layered architecture, conceived to fuse LEO signal strength with favourable geometry while inheriting MEO timing robustness through complementary orbit filtering, promises resilient all-weather global positioning. In addition, GEO satellites maintain a geostationary orbit relative to the Earth’s surface. Theoretically, a single GEO satellite can cover one-third of the Earth’s area, which makes it suitable for static or quasi-static positioning scenarios such as mapping, agriculture, and other related applications. However, it exhibits considerable signal attenuation and a relatively sluggish response speed for dynamic positioning.

\subsection{Fundamentals of INAC}%200-300 292

To enhance the cost-effectiveness of single-function satellite communication or navigation, INAC performed at the satellite network becomes a feasible solution. MEO satellites originally deployed for navigation purposes have been endowed with communication capabilities. For the hardware implementation of INAC systems, the navigation and communication functional modules share the core radio frequency (RF) front-ends and antenna equipment, which does not significantly increase the hardware complexity of satellites, conforming to the lightweight design requirements of satellite payloads. Meanwhile, LEO satellites, primarily designed for communication, have started incorporating navigation data alongside their communication signals.

\begin{figure}[t!]
\centering
\includegraphics[width =0.9\linewidth]{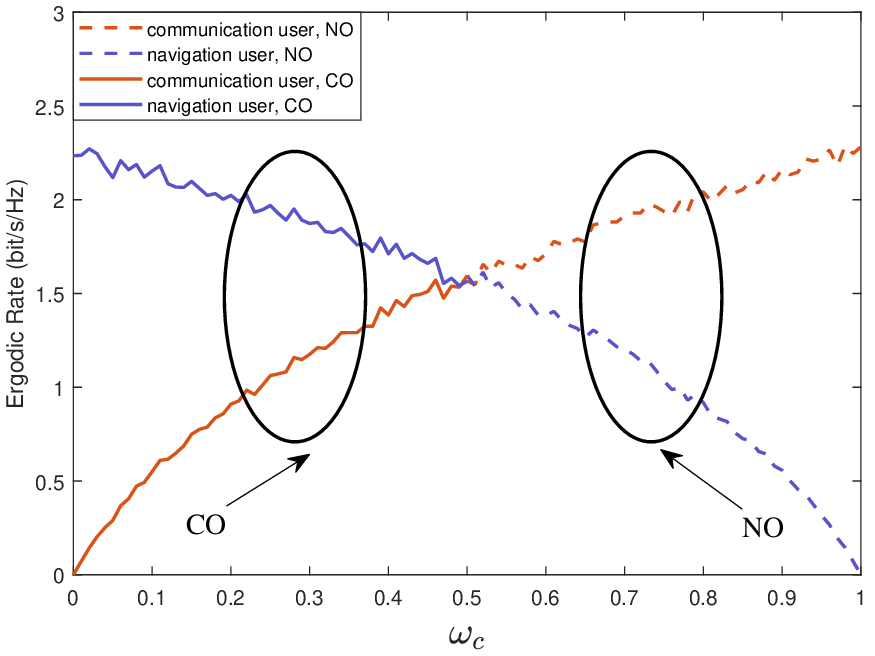}
\caption{\centering Variation of ergodic rate with communication power allocation factors.}
\label{Rvsw}
\end{figure}

In the implementation of INAC systems, encoding schemes are crucial for ensuring robust signal transmission and high data integrity. Navigation signals focus on anti-jamming through pseudo-random noise (PRN) codes, whereas communication signals prioritize high data rates and spectrum efficiency, representing a trade-off between navigation and communication. It is also important to investigate how to merge both signals. The techniques utilized in the ground network aim to place navigation and communication signals on orthogonal communication resources, such as orthogonal frequency division multiplexing (OFDM) and others. Given that, a method for merging communication and navigation signals in the power domain has been proposed in \cite{INAC}. This method is built on two core technologies, which are superposition coding (SC) and successive interference cancellation (SIC). In the downlink channel of the satellite, the INAC signal can be given by
\begin{equation}\label{signal model}
s\left( t \right) = {\omega _N}{s_N}\left( t \right) + {\omega _C}{s_C}\left( t \right),
\end{equation}
where ${\omega _N}$ denotes the navigation power allocation factors. ${\omega _C}$ denotes the communication power allocation factors, where $\omega _N^2 + \omega _C^2 = 1$. ${s_N}(t)$ denotes the navigation signal. ${s_C}(t)$ denotes the communication signal. By adjusting the power allocation factors, satellites can provide corresponding services according to the different demands of users. Then, users can be classified into communication or navigation users based on their requirements. The variation of ergodic rate with communication power allocation factors ${\omega _c}$ of communication and navigation users is shown in Fig. \ref{Rvsw}. For communication signals, the ergodic rate increases as $\omega_c$ rises. Conversely, the ergodic rate of navigation signals shows an opposite trend, which reflects a thought of trade-off based on application scenarios. Furthermore, the ergodic rate of communication/navigation signals is lower than the other, which is due to the use of SIC.

\renewcommand{\arraystretch}{1.0}
\setlength\tabcolsep{2.0pt}%调列距

\section{New Frontier of INAC Networks}

\subsection{LEO-aided INAC Networks}
LEO-aided INAC networks attract growing interest by enhancing both navigation accuracy and communication capabilities.
Existing research mainly targets LEO positioning, LEO-ISLs, and INAC signals for faster and more reliable service.

Rapid progress occurs in Doppler and signals of opportunity (SoP)-based positioning methods. LEO satellites produce significant Doppler shifts due to low altitudes and high speeds, which can be used for velocity and position estimation.
Doppler observation equations, formed from relative velocity and observed frequency shifts, are solved by using least squares or Kalman filtering to estimate receiver velocity and position. However, Doppler positioning does not perform well where satellite signals are weak or unavailable, typically achieving 30 to 50 meters accuracy~\cite{YI}. SoP positioning utilizes existing timing, frequency, and modulation features from non-navigation signals to build mathematical models of receiver location. Measurements from SoP undergo correlation detection, least squares, or filtering to estimate position and velocity. Current SoP method achieves meter-level accuracy in complex environments, serving as an important supplement or alternative to traditional navigation systems~\cite{SoP}.

By enabling satellites to exchange timing signals, ISLs ensure sub-nanosecond synchronization of onboard clocks, thereby playing a crucial role within LEO constellations. Although reliant on GNSS for initial timing, inter-satellite sync maintains the stable time base for precise signals and ranging. To further improve the system resilience, exploring laser-based ISLs to increase data rates is expected. Building upon the precise timing foundation, a major innovation in LEO-INAC involves INAC signal fusion. Kozhaya et al. incorporate navigation codes into communication waveforms similar to 5G signals in the OneWeb constellation, which conserves radio spectrum so that devices can obtain location information even from weak signals, thereby supporting a wider range of applications \cite{Oneweb}. In addition to signal integration, the dynamic geometry of LEO satellites significantly accelerates the initialization of high-precision positioning for ground receivers. While traditional GNSS precise point positioning (PPP) methods may require 30 minutes or longer to converge due to clock and atmospheric error corrections, LEO-based corrections reduce initialization time to under two minutes, allowing receivers to achieve sub-meter accuracy rapidly~\cite{LEOPNT}.

\begin{figure}[t!]
\centering
\includegraphics[width =0.9\columnwidth]{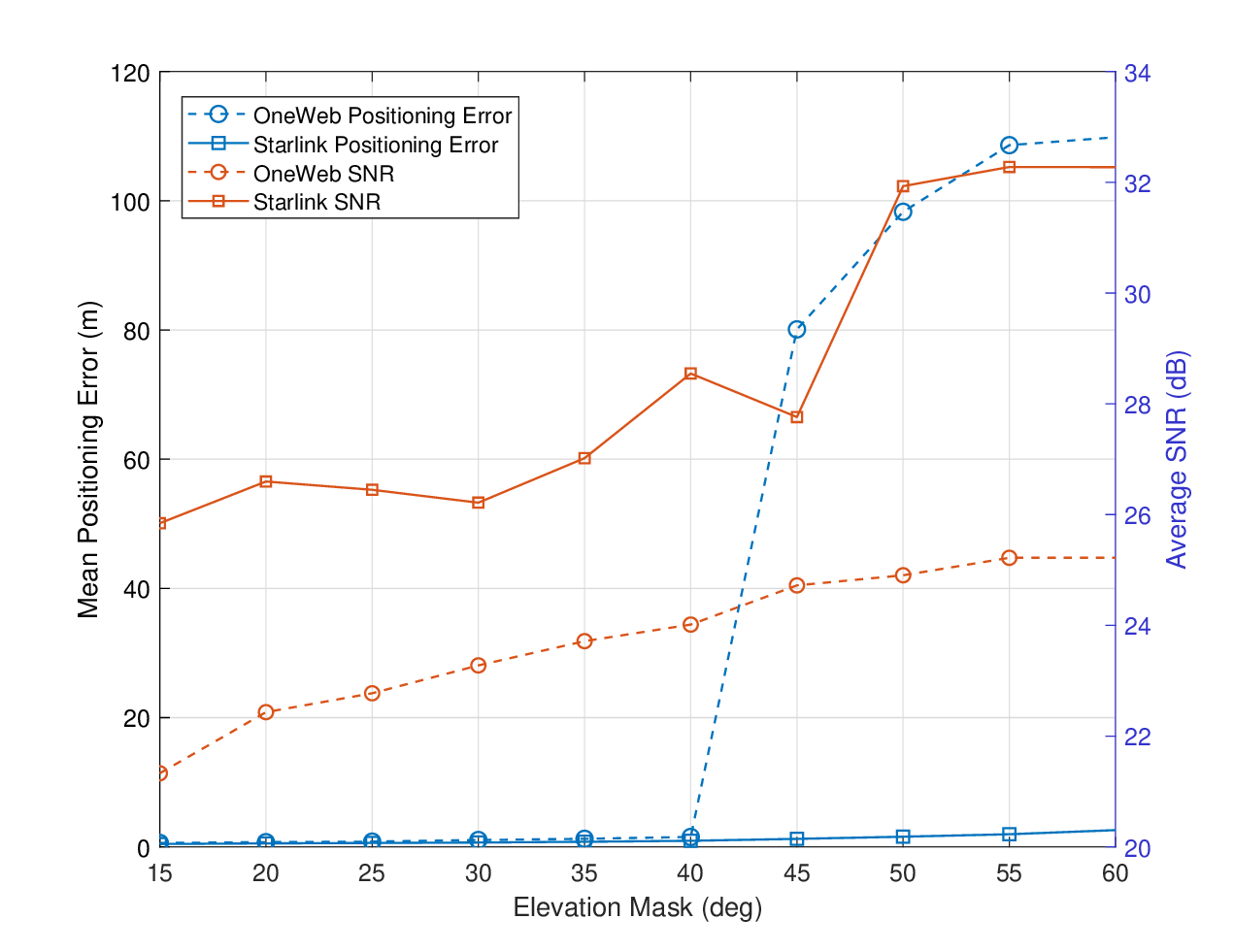}
\caption{\centering Variation of positioning performance and signal quality of OneWeb and Starlink with elevation mask angle.
}
\label{oneweb_star}
\end{figure}

Fig.~\ref{oneweb_star} illustrates the variation in positioning accuracy and average SNR of the OneWeb and Starlink constellations as a function of elevation mask angle. The satellite positions were derived from TLE data at 08:00 UTC on March 26, 2025, with the user fixed at Beijing Jiaotong University. Pseudorange observations were generated from geometric user-satellite distances with additive Gaussian noise of 2~m standard deviation. The position was estimated by using the LS method. Since the simulation was to provide a comparative geometry-based evaluation of constellation robustness, the clock and orbit errors were not modeled separately. Only free-space path loss was considered in the SNR evaluation. 
In the simulation, different elevation cutoff angles were adopted to emulate different obstruction conditions, where lower elevation masks represent relatively open visibility conditions and higher elevation masks correspond to more heavily obstructed environments. For the OneWeb constellation, the number of visible satellites ranges from 97 to 4, while the positioning error remains below 1.5 meters when the elevation cutoff angle is below 40$^\circ$. However, once the cutoff angle exceeds 40$^\circ$, the number of visible satellites drops sharply, resulting in positioning errors increasing to the order of hundreds of meters. The Starlink constellation provides 157 to 19 visible satellites across the elevation mask range. Even at high elevation angles (60$^\circ$), it maintains a positioning accuracy within 2.5 meters. The results indicate that large-scale LEO constellations can provide better positioning robustness under more restrictive visibility conditions.

\subsection{MEO-aided INAC Networks}%312

MEO satellites utilize different resources in navigation and communication. However, the INAC system can significantly enhance the utilization of spectrum and time slot resources, which is compliant with the concept of green communication. For MEO satellites, the vital challenges are the interference and path loss under extremely long communication distances.

From a coding perspective, cyclic code shift keying enhances the anti-interference capability of the INAC signal by periodically applying cyclic shifts to the pseudo-random code, which mitigates multi-path interference, leading to improved SNR performance and greater system stability. In terms of phase and power allocation, phase-optimized constant envelope transmission (POCET) optimizes the phase and power configuration of INAC signals, reducing nonlinear distortion and improving the performance of high-power amplifiers. However, while these technologies primarily enhance anti-interference capabilities and signal transmission stability, they do not yet optimize spectral and energy efficiency. This highlights a key distinction between MEO and LEO systems in terms of received signal power. Consequently, MEO systems require a stronger focus on optimizing communication performance, particularly in terms of minimizing energy attenuation.

\begin{figure}[t!]
\centering
\includegraphics[width =0.9\columnwidth]{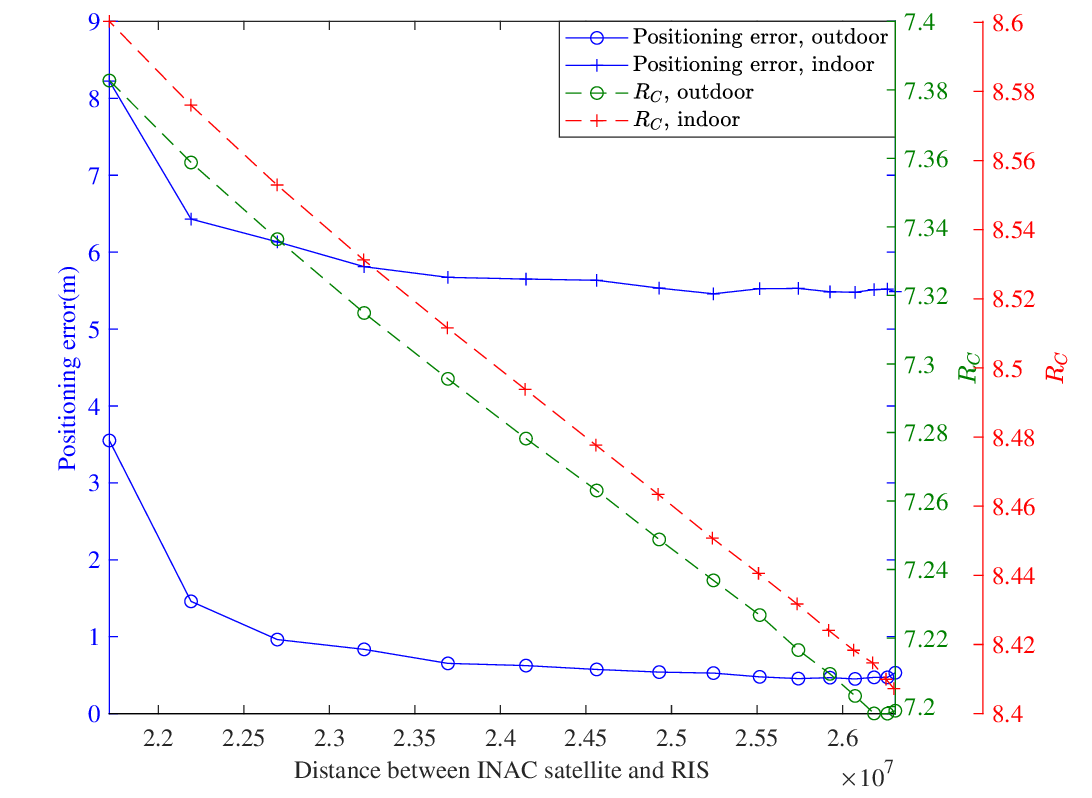}
\caption{\centering Variations in positioning error and ergodic rate versus the INAC satellite-RIS distance.}
\label{NandC}
\end{figure}

However, aforementioned methods primarily focus on interference reduction, without effectively addressing the issue of low received signal power. As the resolution, relay satellites or relay networks work by receiving and retransmitting signals, while high-gain ground stations use directional antennas to concentrate on key signal paths, amplifying signal strength and improving communication reliability. All of them introduce new devices for navigation only and significantly increase the complexity of the system. Fortunately, as a key technology of 6G, reconfigurable intelligent surface (RIS) can dynamically adjust the amplitude and phase of signals, enhancing signal strength, reducing interference and extending LoS (ELoS) links without increasing the complexity of the system significantly~\cite{INAC3}. The employed RIS is a device composed of low-cost passive diodes with a simple planar structure, avoiding complex RF chains. Meanwhile, satellite-based INAC systems do not need explicit channel estimation for either navigation or communication. The positioning accuracy of such systems is inherently immune to imperfect channel state information (CSI). For high-speed LEO scenarios, the INAC signal is first acquired via spread-spectrum processing. SIC is then utilized to separate signals with no loss in navigation precision. Residual interference is suppressed via coding, while active RIS is introduced to enhance signal power. In this article, we simply introduce the channel capacity of communication/navigation signal in RIS-aided NO/CO case as follows~\cite{INAC2}
\begin{equation}\label{channel_capacity}
C_{C/N}^{CO/NO} = B{\log _2}\left( {1 + \frac{{{{\left| {{{\bf{h}}_u}{\bf{\Theta }}{{\bf{H}}_u}} \right|}^2}\omega _{C/N}^2}}{{\delta {{\left| {{{\bf{h}}_u}{\bf{\Theta }}{{\bf{H}}_u}} \right|}^2}\omega _{N/C}^2 + {\sigma ^2}}}} \right),
\end{equation}
where $B$ denotes the bandwidth. ${{\bf{h}}_u}$ denotes the channel vector between RIS and user. ${{\bf{H}}_u}$ denotes the channel vector between RIS and user. ${\bf{\Theta }}$ denotes the RIS reflect matrix. ${\sigma ^2}$ denotes the power of noise. $\delta$ is a binary variable, which takes a value of 1 when calculating the communication signal in NO case and the navigation signal in CO case, and 0 in other cases. In Fig.~\ref{NandC}, with the INAC satellite-RIS distance increasing, the position dilution of precision (PDoP) decreases, which results in a lower positioning error. As distance increases, the positioning error reaches a performance floor, while the achievable ergodic rate declines. This observation also demonstrates the trade-off between positioning and communication performance in INAC system. It should be noted that the above analysis adopts ideal assumptions including perfect CSI and synchronization for performance trend verification. Under imperfect CSI, the channel capacity only degrades slightly, while the positioning accuracy remains unaffected. Accordingly, the performance advantages of INAC systems can still be maintained.

\begin{figure}[t!]
\centering
\includegraphics[width =0.9\columnwidth]{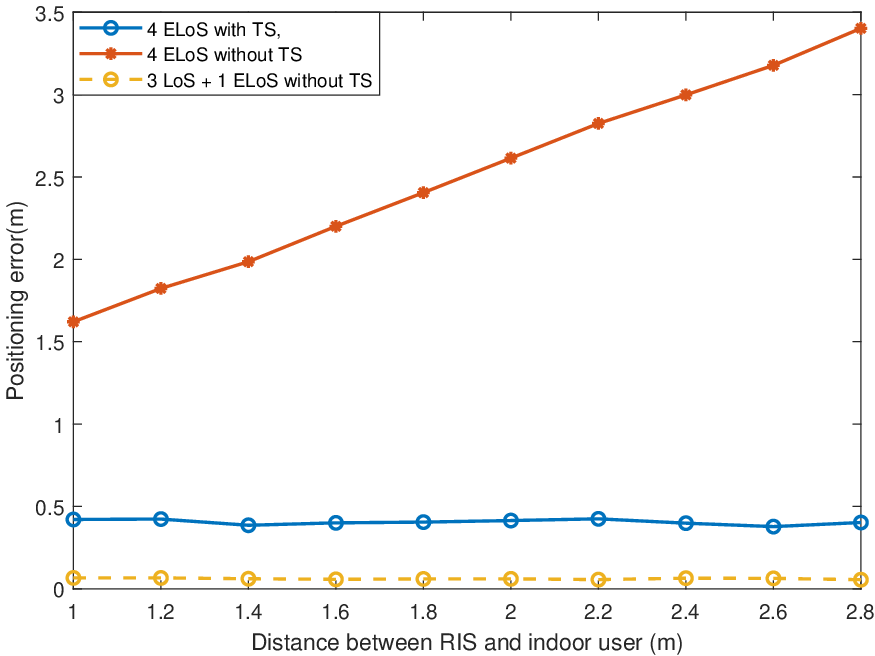}
\caption{\centering Variation of positioning error versus the RIS-user distance.}
\label{four}
\end{figure}

Recent studies have demonstrated the effectiveness of RIS-aided satellite navigation \cite{Satellite_Selection}, highlighting significantly higher positioning accuracy in outdoor environments compared to indoor scenarios. The disparity is largely attributed to the range between RIS and users, which critically influences positioning performance. In addition, the above-mentioned methods require prior knowledge of RIS coordinates. To mitigate these limitations, an active RIS-aided indoor positioning framework with global time synchronization (TS) has been proposed in \cite{ASTAR_Indoor}, which aims to minimize indoor positioning errors and has shown the capability to correct the inaccuracies associated with LSM. Following \cite{ASTAR_Indoor}, a 10~ns synchronization error is adopted in the simulation, corresponding to an approximately 3~m Gaussian ranging error in the observation domain. Fig.~\ref{four} shows the variation of positioning error versus the RIS-user distance, whose result is intended to illustrate the impact of synchronization support on indoor positioning performance.

To better illustrate the differences and characteristics of existing systems in a more intuitive manner, we summarize the representative related works as shown below.
\begin{table*}
COMPARISON OF REPRESENTATIVE SNAC SYSTEMS
\centering
\begin{tabular}{|>{\centering\arraybackslash}m{0.08\textwidth}|
                >{\centering\arraybackslash}m{0.05\textwidth}|
                >{\centering\arraybackslash}m{0.25\textwidth}|
                >{\centering\arraybackslash}m{0.25\textwidth}|
                >{\centering\arraybackslash}m{0.25\textwidth}|}
\hline
Reference & Orbit & Key Technologies & System Architecture & Innovation Points \\
\hline
\cite{Xu2025} & LEO & Joint Pseudorange-doppler positioning & Starlink; Multi-constellation LEO & High-sensitivity detection; Orbit error suppression \\
\hline
\cite{INAC} & MEO & NOMA-RIS; INAC & RIS-aided INAC network & Power allocation \\ 
\hline
\cite{YI} & LEO & Phase-Time method; Doppler measurement & Iridium NEXT signal processing & Full-link signal enhancement \\ 
\hline
\cite{INAC3} & MEO & STAR-RIS, INAC, NOMA & STAR-RIS aided INAC & Urban canyon \\ 
\hline
\cite{ASTAR_Indoor} & MEO & ASTARS; Extended path  & ASTARS aided satellite navigation & Distance correction \\ 
\hline
\end{tabular}
\label{table1}
\end{table*}
Based on the above, a pivotal deep insight is LEO-MEO’s mutually reinforcing integration. MEO provides a stable global PNT backbone, while LEO delivers low-latency, high-signal-strength performance. On the one hand, RIS helps mitigate the path loss in MEO systems. On the other hand, RIS is also capable of extending the coverage of LEO systems.

\section{The Way Forward}

\begin{figure}[t!]
\centering
\includegraphics[width =0.9\columnwidth]{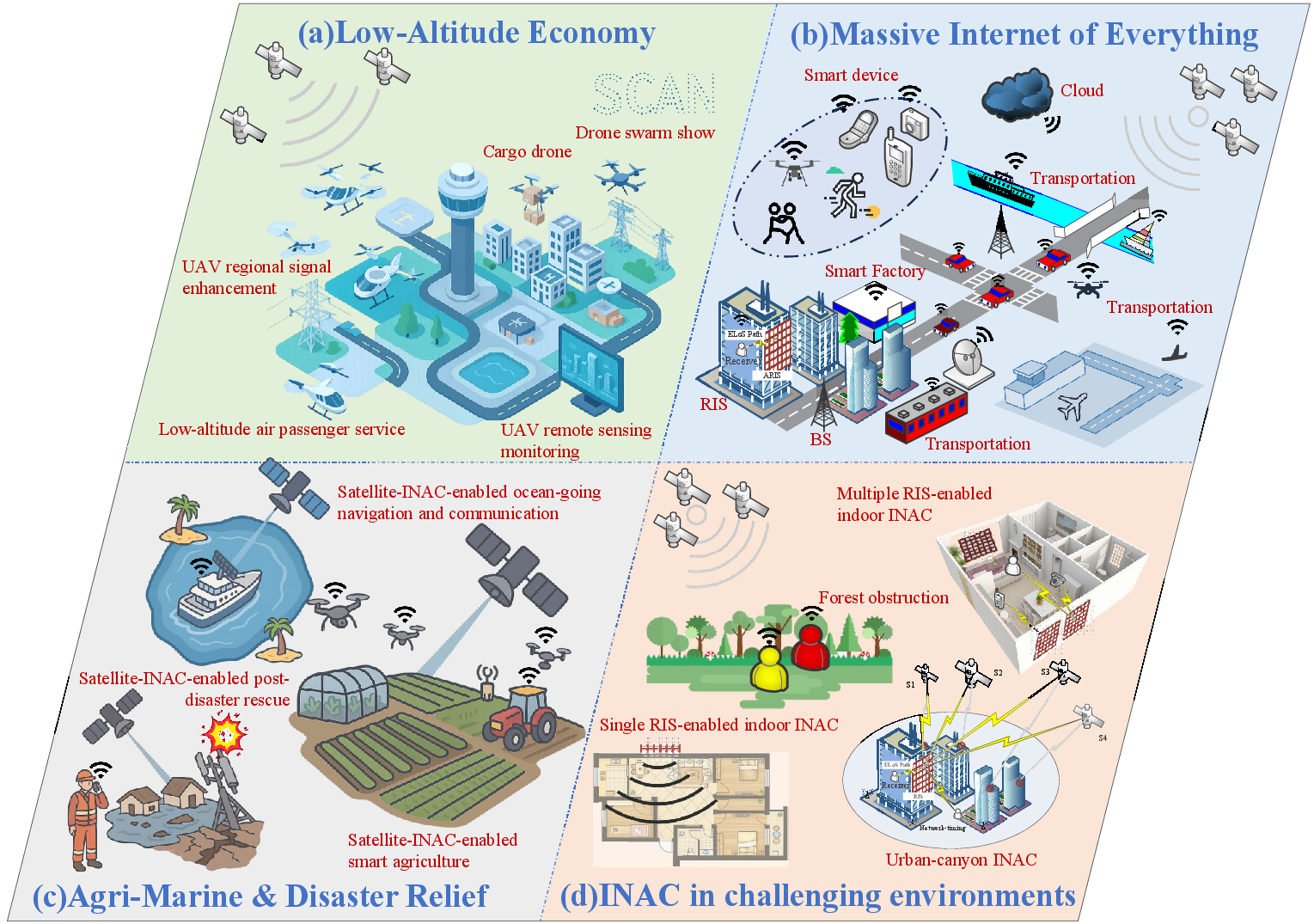}
\caption{The potential applications of satellite INAC networks.
}
\label{app}
\end{figure}

Leveraging seamless air-space-ground-sea coverage and the native integration of communication and positioning services, the satellite INAC is regarded as a key capability-enhancement layer within the holistic 6G architecture. Fig.~\ref{app} illustrates INAC’s wide-ranging potential. Nevertheless, fully unlocking this potential requires sustained research efforts aimed at simultaneously expanding channel capacity and achieving better positioning accuracy.

\subsection{Higher Channel Capacity}

With the rapid advancement of 6G, achieving higher communication rates in satellite services is fundamental to realizing ubiquitous communication. To enhance communication performance, technologies such as unmanned aerial vehicles (UAV), RIS and others can be integrated into satellite INAC networks. The evolving demands have introduced new directions for future INAC research, including efficient space-based computing, unified resource management for satellite systems, intelligent spectrum sharing between space and ground, simplified access and handover processes, multi-beam communication technology, grant-free access technologies, and novel coding techniques. While enhancing the communication rate of INAC systems presents numerous challenges, it also opens up exciting opportunities for further innovation.

\textbf{New Challenges:}
\begin{itemize}
\item Limited spectrum resources: The rapid growth of satellite systems has exacerbated the scarcity of spectrum resources. To address this, dynamic spectrum sharing is a promising solution, which enables secondary users to access unused licensed spectrum while limiting interference to primary users. Investigating the performance analysis and optimization strategies for spectrum sharing technologies presents an emerging challenge.
\item Low energy efficiency: Due to the limited power resources, satellite systems urgently need to enhance energy efficiency through novel multi-access techniques. SC and SIC allow multiple users to share the same power resource block by superimposing their signals in the power domain, thereby improving energy efficiency. Investigating the appropriate user allocation strategy in a dynamic satellite network represents one of the future research challenges.
\item Diverse demands of users: In the 6G satellite system, users with different communication rate requirements are located at diverse locations. Multi-beam technology can be investigated to better manage diverse user requirements, thereby enhancing overall system capacity and accommodating different user types. Allocating distinct beams to meet diverse user requirements stands as a challenge for future research.
\item Limited hardware architecture: The hardware on satellites is highly constrained by size, weight, and power, which severely restricts the complexity of onboard hardware. To achieve higher channel capacity, developing more efficient hardware to maximize cost efficiency is crucial in future research.
\end{itemize}

\textbf{New Opportunities:} INAC presents numerous opportunities across a variety of emerging applications. Leveraging the global coverage of satellites, INAC-enabled systems can provide reliable positioning and communication support for low-altitude economy applications, such as UAV performances and delivery. Moreover, INAC plays a vital role in enabling intelligent agriculture, maritime operations, and emergency response scenarios, where precise positioning and robust communication links are essential. In such domains, maintaining connectivity under dynamic and resource-constrained conditions is expected.

\subsection{Better Positioning Accuracy}

Improving satellite navigation accuracy necessitates theoretical advancements across signal design, space segment, ground segment, and user segment.

\textbf{1) Signal Design}.
For MEO satellites, efforts emphasize multi-signal fusion and multi-frequency processing to boost interference resistance and positioning precision, guided by information-theoretic methods for mitigating multi-path effects. For LEO satellites, advanced navigation and signal-fusion mechanisms are expected, including joint modulation schemes, time-frequency resource sharing, and combined positioning algorithms to enhance efficiency and reliability in complex environments. Security receives increased attention through embedding low-power cryptographic authentication within navigation signals to protect against spoofing, while ensuring trustworthy positioning services.

\textbf{2) Space Segment.}
Future research in the MEO segment aims to develop analytical models for ultra-stable optical atomic clocks, with parallel efforts devoted to building sparse inter-satellite laser-synchronization networks. In the LEO segment, efforts concentrate on designing compact cold-atom and miniature rubidium clocks, whose clock drift under orbital perturbations needs careful evaluation. Advanced autonomous precise orbit determination methods are potentials, which incorporate refined perturbation models, including solar radiation pressure, atmospheric drag, gravitational asymmetry, and attitude dynamics.

\textbf{3) Ground Segment.}
Future research in the ground segment focuses on advanced atmospheric delay modeling, faster PPP convergence, and network-assisted time synchronization. Techniques like low-dimensional stochastic field models, including Karhunen--Lo\`eve expansions and sparse Gaussian processes, efficiently represent ionospheric and tropospheric errors to simplify real-time corrections. Bayesian data assimilation frameworks combining GNSS data,
LEO atmospheric occultation, and numerical weather models aim to improve accuracy, while reducing convergence time. Additionally, random matrix theory and convex optimization guide optimal ground station placement by quantifying the relationship between station geometry, correction precision, and overall robustness. Network-assisted time synchronization can provide a critical foundation for delay-based indoor positioning. Accurate timing information is essential to indoor satellite-based ranging and positioning, which remains challenging in practical indoor environments.

\textbf{4) User Segment.}
Future research in the user segment focuses on developing theoretical frameworks for multi-source data processing, which employ factor graphs and Bayesian inference to integrate MEO pseudo-range, LEO Doppler and carrier-phase measurements, inertial sensors, and terrestrial synchronization signals into unified probabilistic models. More precise indoor satellite-positioning algorithms and techniques for enhancing accuracy in urban canyons are likewise highly anticipated.
Additionally, advanced satellite selection techniques based on information geometry optimize real-time satellite subset choice.

\subsection{AI-Enabled Satellite INAC}

Beyond the widely adopted OFDM waveform employed in 4G and 5G systems, several alternative waveforms have recently emerged as promising candidates for satellite communications. Among these, the orthogonal time frequency space (OTFS) waveform~\cite{OTFS}, which operates in both the delay and Doppler domains, has attracted significant attention. Such novel waveform designs are potentially critical for fully utilizing the joint capabilities of delay and Doppler dimensions for INAC services. For instance, in MEO satellite navigation, the Doppler effect is typically considered an invalid parameter for positioning. In contrast, INAC systems leverage the coupling between delay and Doppler domains.

In addition, the uni-cast and multi-cast service requirements in satellite communication and navigation inherently conflict with one another. In satellite navigation, multi-cast signals are essential, as the same signal must be broadcast to all users within the coverage area. In contrast, satellite communication often relies on uni-cast transmission, where dedicated signals are allocated to individual users. The divergence presents a challenge in designing unified signal waveforms. Nevertheless, OFDM offers a practical solution by enabling flexible resource allocation to support both uni-cast and multi-cast transmissions simultaneously. However, effectively harnessing time, frequency, code, power, delay, and Doppler resources across both LEO and MEO satellite systems remains a complex yet critical challenge for the realization of AI-enabled INAC architectures.

\section{Conclusions}

This article elucidated the foundational principles, new frontiers and future prospects of satellite INAC systems. The navigation and communication functions within both LEO and MEO satellite systems were made possible by INAC. However, fully unlocking the potential of INAC systems depends on addressing practical challenges related to channel capacity, positioning accuracy, and AI-driven design, necessitating continued research and development.

\bibliographystyle{IEEEtran}
\bibliography{IEEEabrv,INAC_Com_Mag}

\begin{IEEEbiography}[{\includegraphics[width=1in,height=1.25in,clip,keepaspectratio]{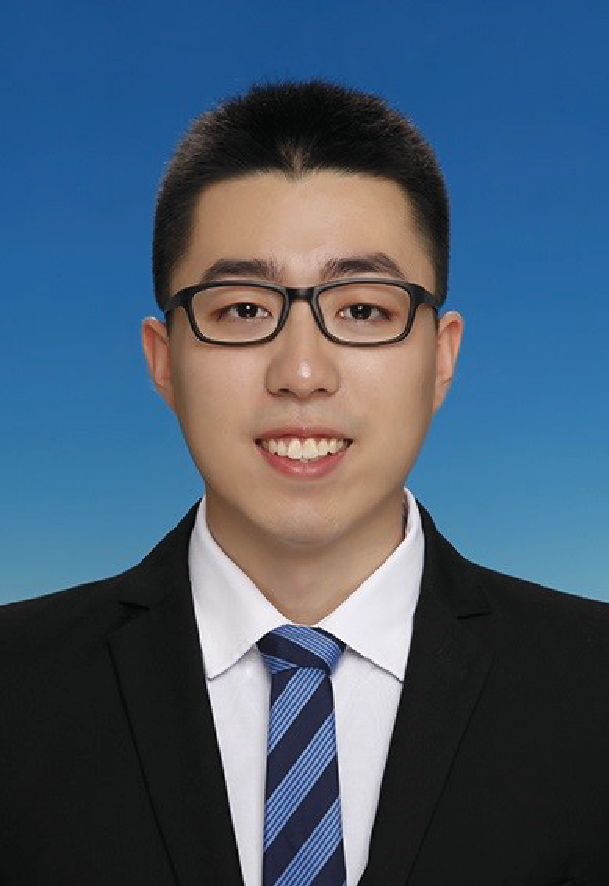}}]{Tianwei Hou (Member, IEEE)}
received the Ph.D. degree from Beijing Jiaotong University (BJTU) in 2021. He was a Visiting Scholar with the Queen Mary University of London (QMUL) from September 2018 to November 2020. Since 2021, he has been an Associate Professor with BJTU. His current research interests include next generation multiple access (NGMA), reconfigurable intelligent surface (RIS) aided communications, UAV communications, multiuser multiple-input multiple-output (MIMO) communications, and stochastic geometry. He has served as a TPC Member for many IEEE conferences, such as GLOBECOM, and VTC. He served as the Publicity Officer for the Next Generation Multiple Access Emerging Technology Initiative (NGMA-ETI). He has granted a Marie Skłodowska-Curie Fellowship by European Research Executive Agency in 2023. He has been selected as a Young Elite Scientist Sponsorship Program by China Association for Science and Technology in 2022. He received the Exemplary Reviewer of {\sc IEEE Communication Letters} and {\sc IEEE Transactions on Communications} in 2018, 2019, and 2022. He has served as the Co-Chair for the 2nd, 4th, and 5th NGMA-for-Future-Wireless-Communication Workshops in IEEE VTC 2022-Fall, IEEE VTC 2023-Spring, IEEE ISCT-2022, and IEEE PIMRC 2023. He has also served as the Co-Chair for the RIS and Smart Environments Symposium of IEEE ICCT 2023. He serves as the leading Guest Editor for IEEE IoT-J special issue on Next Generation Multiple Access for Internet of Things.
\end{IEEEbiography}

\begin{IEEEbiography}[{\includegraphics[width=1in,height=1.25in,clip,keepaspectratio]{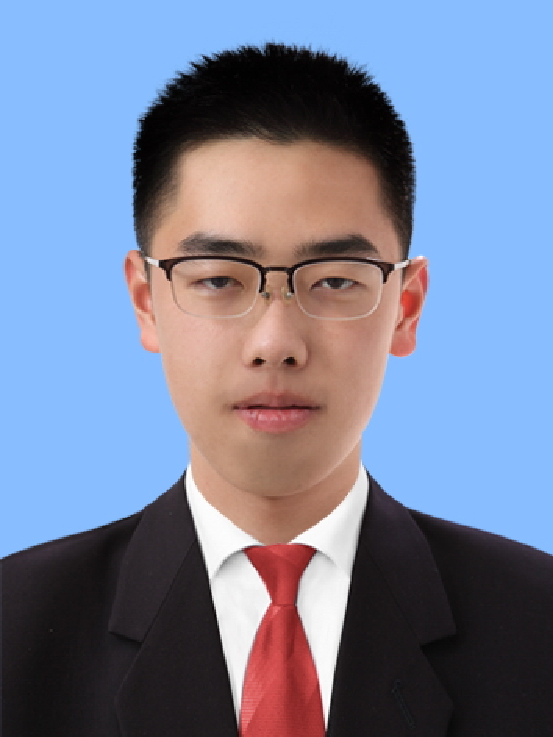}}]{Da Guan}
received the B.S. degree in Communication Engineering from Beijing Jiaotong University (BJTU), Beijing, China, in 2023, where he is currently pursuing the Ph.D. degree.
His research interests include satellite-integrated navigation and communication, reconfigurable intelligent surface (RIS)-aided communications, and near-field sensing. He was a recipient of the Best Paper Award at the \textit{IEEE International Conference on Information Systems and Computing Technology (ISCTech)}, 2022. He has served as a TPC Member for ICC 2026 WS-29 - 6GSatComNet.
\end{IEEEbiography}

\begin{IEEEbiography}[{\includegraphics[width=1in,height=1.25in,clip,keepaspectratio]{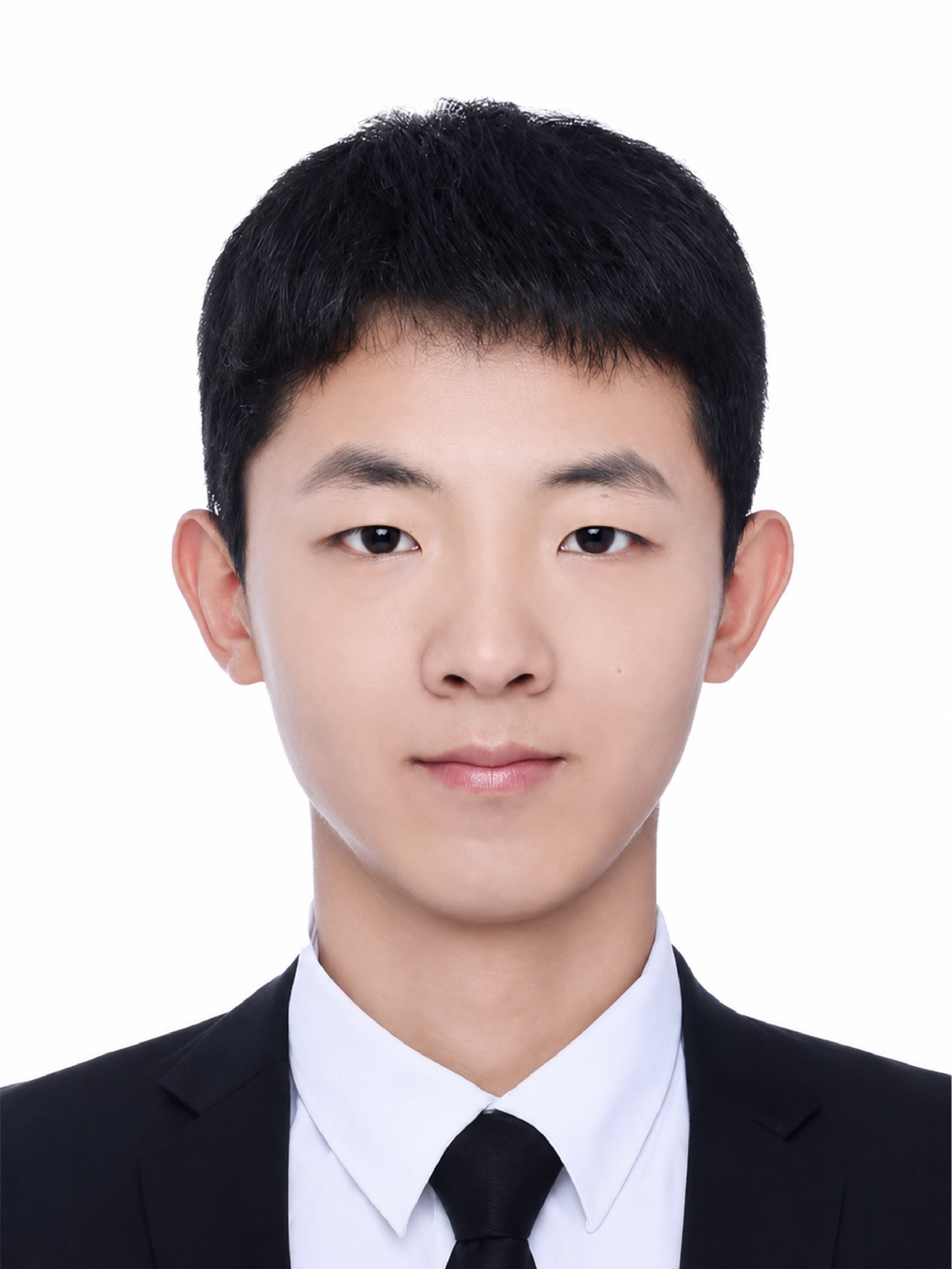}}]{Yu Zhang}
received the B.S. degree from Beijing Jiaotong University (BJTU), China, in 2025. He is currently pursuing the Ph.D. degree with the School of Electronic and Information Engineering, Beijing Jiaotong University.
His research interests include low Earth orbit (LEO) satellite positioning and navigation technologies, and integrated navigation and communication (INAC) technologies.
\end{IEEEbiography}

\begin{IEEEbiography}[{\includegraphics[bb=0 0 500 600,width=1in,height=1.25in,keepaspectratio]{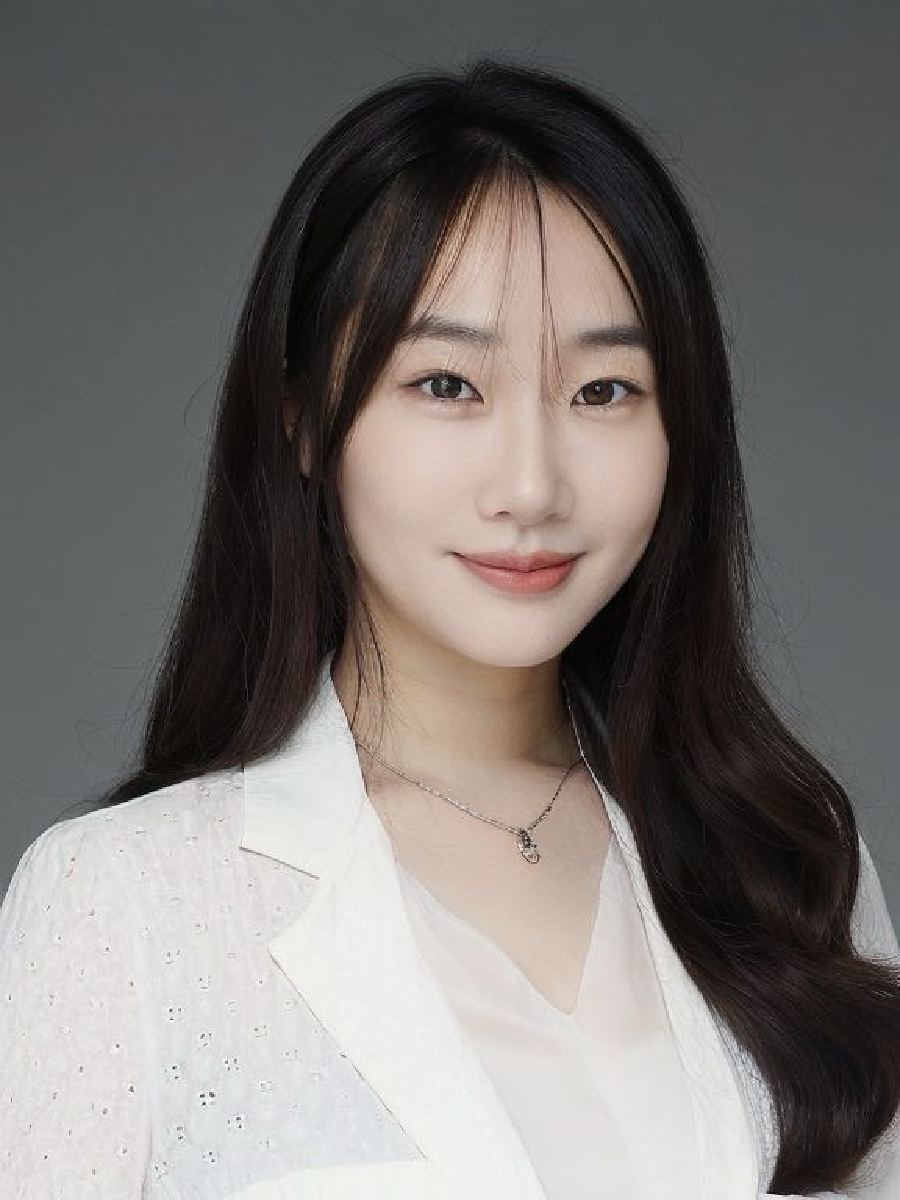}}]{Anna Li (Member, IEEE)}
received the Ph.D. degree in computer science from the School of Electronic Engineering and Computer Science, Queen Mary University of London, London, U.K., in 2023. She has been a Lecturer with the School of Computing and Communications, Lancaster University, Lancaster, U.K., since September 2023. She is also a Marie Curie Fellow at Katholieke Universiteit Leuven (KU Leuven), Leuven, Belgium. Her research expertise encompasses wireless sensing, satellite navigation and communication systems, signal processing, and deep learning.
\end{IEEEbiography}

\begin{IEEEbiography}[{\includegraphics[width=1in,height=1.25in,clip,keepaspectratio]{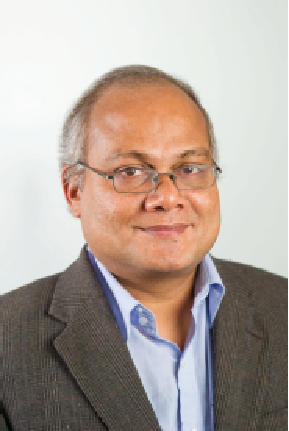}}]{Arumugam Nallanathan (S'97-M'00-SM'05-F'17, Fellow, IEEE)}
received the B.Eng. degree (Hons.) in electrical and electronic engineering from the University of Peradeniya, Sri Lanka, in 1991, the CPGS degree in electrical and electronic engineering from the University of Cambridge, Cambridge, U.K., in 1994, and the Ph.D. degree in electrical and electronic engineering from The University of Hong Kong, Hong Kong, in 2000.

He was an Assistant Professor with the Department of Electrical and Computer Engineering, National University of Singapore, Singapore, from August 2000 to December 2007. He was with the Department of Informatics, King's College London, London, from December 2007 to August 2017, where he was a Professor of wireless communications from April 2013 to August 2017, and a Visiting Professor from September 2017 to August 2020. He has been a Professor of wireless communications and the Founding Head of the Communication Systems Research (CSR) Group with the School of Electronic Engineering and Computer Science, Queen Mary University of London, London, U.K., since September 2017. He has published more than 700 technical papers in scientific journals and international conferences. His research interests include artificial intelligence for wireless systems, beyond 5G wireless networks, and the Internet of Things.

Dr. Nallanathan was a co-recipient of the Best Paper Awards presented at the IEEE International Conference on Communications 2016 (ICC'2016), IEEE Global Communications Conference 2017 (GLOBECOM'2017), and IEEE Vehicular Technology Conference 2018 (VTC'2018). He was also a co-recipient of IEEE Communications Society Leonard G. Abraham Prize in 2022. He has been selected as a Web of Science Highly Cited Researcher in 2016, 2022-2024. He received the IEEE Communications Society SPCE Outstanding Service Award 2012 and IEEE Communications Society RCC Outstanding Service Award 2014. He was a Senior Editor for {\sc IEEE Wireless Communications Letters}, an Editor for {\sc IEEE Transactions on Wireless Communications}, {\sc IEEE Transactions on Communications}, {\sc IEEE Transactions on Vehicular Technology}, and {\sc IEEE Signal Processing Letters}. He served as a Guest Editor for numerous special issues of {\sc IEEE Journal on Selected Areas in Communications} (JSAC). He served as the Chair for the Signal Processing and Communication Electronics Technical Committee of IEEE Communications Society and a technical program chair and a member of technical program committees for numerous IEEE conferences. He is an IEEE Distinguished Lecturer.
\end{IEEEbiography}

\end{document}